\documentclass[preprint,12pt]{elsarticle}
\usepackage[utf8]{inputenc}
\usepackage{cancel}
\usepackage{graphicx}
\usepackage{epsfig}
\usepackage{url}
\usepackage{epstopdf}% To incorporate .eps illustrations using PDFLaTeX, etc.
\usepackage[caption=false]{subfig}% Support for small, `sub' figures 
\usepackage[american]{babel}

\usepackage{amssymb}
\usepackage{amsmath}
\usepackage{graphicx}
\usepackage{rotating, lscape, longtable, tabu, booktabs, siunitx, multirow}
\usepackage[capitalise, noabbrev]{cleveref}
\usepackage{verbatim}
\usepackage{mathtools}

\begin{document}

\begin{frontmatter}

\title{Application of Benford-Newcomb Law with Base Change to Electoral Fraud Detection.}

\author{Eduardo Gueron\corref{cor1}}
\ead{eduardo.gueron@ufabc.edu.br}

\author{Jerônimo Pellegrini}
\ead{jeronimo.pellegrini@ufabc.edu.br}

\address{Centro de Matemática Computação e Cognição - Universidade Federal do ABC}

\cortext[cor1]{Corresponding author}

\begin{abstract}

The invariance of Benford-Newcomb law under base changing is employed to test whether or not some data follow such distribution. Taking into account the Brazilian senate election in 1994, changes in the numerical base were able to evidence probable fraud. 

\end{abstract}


\end{frontmatter}
%\title{Application of Benford-Newcomb Law with Base Change to %Electoral Fraud Detection.}

%\author{Eduardo Gueron}

%\fnref{\ead email@email}%\corref{cor1}\ead{eduardo.gueron@ufabc.edu.br}}%~\textsuperscript{a}\thanks{CONTACT E.~G. Email: eduardo.gueron@ufabc.edu.br} and Jerônimo Pellegrini~\textsuperscript{b}\thanks{CONTACT J.~P. Email: jeronimo.pellegrini@ufabc.edu.br}

%\affiliation{Universidade Federal do ABC}


%\affiliation{organization={},%Department and Organization
 %           city={},
 %           postcode={}, 
 %           state={},
 %           country={}}

%\begin{abstract}

%\end{abstract}



%\maketitle

%\begin{keyword}

%

%\end{keyword}

%\end{frontmatter}

\section*{Introduction}

 Simon Newcomb, a Canadian astronomer in the nineteenth century, perspicaciously observed that the first pages of logarithmic tables (corresponding
to lower digits) were much faster than the last ones. He than
concluded that his colleagues were using numbers beginning with the digit
1 much more often than other numbers.
From empirical observation, the Canadian proposed a law for the
frequency of the first two digits for what he called ``Natural Numbers.''
His observations and some heuristic caolculations were published in
1881~\cite{key-2}.

About 50 years later an American physicist, Frank Benford, also observed
the difference in the dirtiness of the first pages and made some numerical experiments using a different data set~\cite{benford}. 

After them,  we name the Benford-Newcomb Law, which predicts that for some sets of numbers, the leading digit $d$ occurs with probability
\[
P(d)=\log\left( 1+\frac{1}{d} \right).
\]
This law generalizes for the following digits and also may be taken on different bases.~\cite{annals}

When large enough accounting data must follow Benford-Newcomb's
distribution, such a law has been presented
as a statistical indicator of fraud.  It has been used to detect fraud in several different sets --
financial statement fraud, tax evasion, among others~\cite{miller,nigrini} and also, with some controversy, as a tool for election monitoring~\cite{iran}.

In this article, we aim to show that  Benford-Newcomb can be employed to point out a possible electoral fraud, particularly when different number bases are considered. To support our statement, we performed a case study of the 1994 Senator election in the State of Bahia, Brazil. On that occasion, there were allegations of fraud but no evidence was found proving it. Nonetheless, no robust statistical analysis was used in the investigation at that moment.

%For instance, we have found strong evidence of illegal %interference in the 1994 Senator election in the State of Bahia, %Brazil. There weore claims of fraud at that time, but no strong %evidence was found.

\section*{Using Base Invariance}

It is known that Benford's law is invariant to scale\cite{scale}, i.e., the distribution does not change under multiplication by a constant, an interesting study also mentions the convergence to NBL under some specific Markovian process\cite{markov}. Moreover,  such a distribution is base invariant, i.e., not affected by a change of base~\cite{Schatte}. T. Hill went even further showing that ``Base invariance implies Benford's Law'' \cite{baseinvariance}. 

Although the results on base invariance are well known, so far as we are aware no method in the literature employs such property to detect any kind of fraud. Thus, we present in this paper alternative techniques capable of quantifying deviations from expected Newcomb-Benford distribution by changing the numerical base of data.  

So we therefore invoke the Newcomb-Benford law for a number written in an arbitrary base $b$. It verses on the probability of finding $d$ as leading digit and can be cast as
\begin{equation}
P(d)=\log_{b}\left( 1+\frac{1}{d} \right),\label{eq:prob}
\end{equation}
or
\[
\frac{\ln \left( 1+\frac{1}{d} \right) }{\ln(b)}.
\]
o
By using other bases, the possibility of studying the same set of data with more parameters is open and therefore a more complete analysis is allowed.  It is worthy mentioning an interesting study about distributions in different bases presented by Frank Benford (grandson)~\cite{base}.

\subsection*{Benford and Elections}

%\subsection*{Controle: Benford and Generic Elections}
To obtain, in some way, quantitative data that indicates the level of fraud in a given election,
different statistical methods have been employed over time. Among them, Benford's first digits law has
become very popular in recent years, being used, for example, to discuss a possible fraud in the Iranian
presidential elections of 2009 \cite{iran}

Nevertheless, the use of Benford-Newcomb is far from fully established and not even well accepted as an electoral system
fraud analysis tool. According to Deckert, Myagkov and Ordeshook \cite{benfordruim}  the lack of a clear
political model and the absence of well-defined patterns have been leading to misjudgments. They also claim
that the comparison of the rigged election with mathematical simulations of fair and unfair
elections was not sufficient to justify whether or not a fair election should follow a Benford-Newcomb distribution
analyzing the first and second digit distribution. Here, fair or unfair elections mean, respectively,
elections with the absence or presence of notorious relevant fraud. 

o
Certainly the question regarding if the votes in fair elections follow a Benford-Newcomb distribution depends on several not very
well-defined factors. Nonetheless, a comparison between similar elections may settle down such a problem. 
In that sense, the work regarding the analysis of the Iranian elections  \cite{iran} compared first-round presidential elections from different countries with geopolitics and voting dynamics quite dissimilar to Iran. For example, voting is mandatory in Brazil, while not in France, so any comparison between those would be questionable but data form both countries election was used in the analysis. 

Therefore we shall compute empirically among a set of similar elections the maximum acceptable deviation from the Benford-Newcomb distribution to have a parameter to declare a possible fraud. Evidently, a safety distance from the average may be in order, and a gray interval for ``not conclusive'' may be defined, although the example in this paper does not need these -- the distance from the average ends up being large enough that a simple inspection suffices.

We thus proceed this way to analyze some Brazilian majority elections from 1994 to 2018. We were particularly interested in the senate elections in the seven states with more municipalities, respectively, Minas Gerais, São Paulo, Rio Grande do Sul, {\bf Bahia}, Paraná, and Goiás. These were chosen due to some previously suspected fraud in Bahia Elections in 1994.

%(citar cidades como zonas eleitorais e ver termo técnico para isso %no contexto da distribuição) 

%{\bf citar livro  do Jeronimo nesse capítulo}
\section*{Bahia Senate Elections}

In the election in 1994, Brazilian voters were to choose among several candidates to represent their states in the Senate.
Each Brazilian state elected its two most voted candidates.

At that time one of the candidates, Antonio Carlos Magalhães (ACM),  dominated Bahia politics and therefore was the favorite to be the most voted. The race for the second place, however, was quite fierce between Waldeck Ornellas and Waldyr Pires -- the first being an ally of ACM. After all, Ornellas beat Pires by a difference of $0,06\%$.

Not surprisingly, unusual results were noticed in some electoral sections -- for instance, the absence of invalid votes in some urns, which were common in elections at the time, since votes were written down on paper ballots and dissatisfied voters would take the opportunity to express their anger directly on the ballot. Among other reports at that time, Janio de Freitas, a journalist at Folha de São Paulo wrote (in Portuguese)``(...){\em The polling maps in Bahia revealed an electoral phenomenon: in more than 1,400 ballot boxes, the candidate for senator Waldeck Ornelas has many more votes than his political godfather Antônio Carlos Magalhães, who sponsored him only to complete the necessary pair of candidates, headed by himself }(...)" \cite{fsp_janio}. Pires turned to the Regional Electoral Court\footnote{Regional courts are responsible for the control of the whole electoral process in their jurisdiction (in this particular case, the state of Bahia). } and claimed fraud, but
his petition was denied~\cite{treba}. The case was also analyzed by higher courts but despite several signs of electoral fraud, the alleged lack of evidence prevented any kind of recall procedure~\cite{fsp, conjur, stf}.

In this section, we review that election using data  extracted  from Brazil's National Electoral
Court\footnote{``TSE'' in Portuguese}~\cite{tse}, and compare it to senator elections in other Brazilian states with approximately as many cities as the state of Bahia. 

The sample considered for this study consisted of the votes for the first three candidates grouped by cities in the states of  Bahia (BA), Paraná (PR), São Paulo (SP), Minas Gerais (MG), Goiás (GO), Rio Grande do Sul (RS) and Santa Catarina (SC) from 1994 to 2018. These states were chosen because Bahia is the 4th state in the number of municipalities and these are the first seven, in decreasing order: MG,~SP,~RS,~BA,~PR,~SC,~GO.  
 
One of the main challenges dwells in the search for parameters that measure how close to Benford-Newcomb should the vote distribution be considered an unsuspected election. There is no unmitigated answer for that, nevertheless, we may argue that a set of similar elections can settle up the standards.  A comparison of the seven selected states above mentioned should give, at least, outliers.

\subsection*{Methods and Results}
In order to analyze the data obtained from the Brazilian elections, different approaches are used to somehow quantify the possibility of electoral fraud by considering that the distribution of votes per section follows the First Digit Benford-Noewcomb Distribution.

Some of the numerical results were obtained using R with some additional packages\cite{packages}

\subsubsection*{Method 1 - Digit Ratio}

 Let us first define a function $R(d_1,d_2)$ that computes the ratio between the probability of finding $d_1$ as a leading digit and the probability of finding $d_2$ as a leading digit, i.e., $$R(d_1,d_2)=\frac{P(d_{1})}{P(d_{2})}.$$

From Eq.\ref{eq:prob} we  have

\begin{equation}
R(d_{1},d_{2})=\frac{\ln(1+\frac{1}{d_{1}})}{\bcancel{\ln(b)}}\frac{\bcancel{\ln(b)}}{\ln(1+\frac{1}{d_{2}})},\label{Procedimento1}
\end{equation}o
which then shows that $R(d_{1},d_{2})$ does not depend on base $b>\max\{ d_1,d_2 \}.$

So first we shall plot $\mbox{Base} \times R(d_1,d_2)$ for the group of states previously chosen. If the data set followed perfectly BND a horizontal straight line would be plotted. Therefore, moving away from the horizontal line might somehow give us an indication of fraud. However, as repeatedly mentioned above, we need to compare all the measured distances for each election to search for outliers.

\begin{figure}[htb!]
\includegraphics[width=11cm,height=7cm]{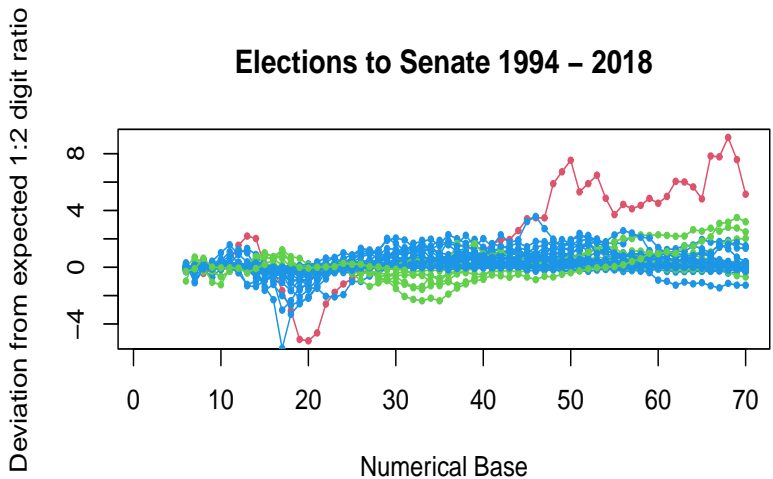}
\caption{Error Function for the ratio $1/2$. We calculate, for each base, $f(d_1,d_2)-f(d_2,d_1)$. The red Line represents Bahia 1994 Senate elections, Green Lines represent other years Bahia senate elections and Blue Lines are the Senate Elections for the other considered states.}\label{grn}
\end{figure}

From the data of the elections above-defined we thus compute the function $R(1,2)$,  given by Eq.\ref{Procedimento1}, for different bases (in this case, from $6$ to $70$) and plotted the results at Fig.~\ref{grn} - the choice for the digits 1 and 2 will be explained later.  It is easy to notice, by simple visual inspection, that the election corresponding to the red-line, Bahia Senate-1994, is an outlier. So, we have the first indication of fraud in that voting.

The figure also helps one see why using only a single basis wouldn't be enough -- for a number of bases, the distribution happened to seem to be within the expected, {\em including base ten}.

\subsubsection*{Method 2 - Kullback Leibler}

We also compute the  Kullback–Leibler divergence to compare the first digit distribution of a sample to the expected first digit proportion in a Benford-Newcomb distribution for different bases.  Actually, we use the so-called Intrinsic Discrepancy Loss, given by the following equation \cite{kullback},
\begin{equation}
    KLD_b=\min\left\{ \sum_{d=1}^{b-1} P_B(d)\log\left(\frac{Q_S(d)}{P_B(d)}\right), 
    \sum_{d=1}^{b-1} Q_S(d)\log\left(\frac{P_B(d)}{Q_S(d)}\right)\right\},
\end{equation}o
where $Q_S(d)$ is the proportion of digit $d$ in the sample written in base $b$ and $P_B(d)$ is the probability of finding leading digit $d$ in the Benford-Necomb distribution written in  $b$, Eq.\ref{eq:prob}.

\begin{figure}[ht]
\epsfig{file=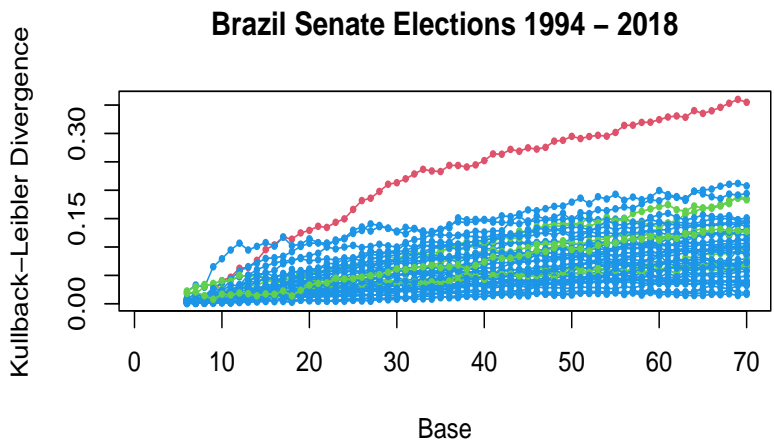,width=11cm,height=7cm}
\caption{Kullback-Leibler Symmetric distance for each base. The red line represents Bahia-1994 election, Green Lines Bahia elections after 1994 and the blue ones are all the other considered senate elections.}\label{intrinsic}
\end{figure}

Results are presented in Fig.\ref{intrinsic}. Again, the red line represents the Bahia-94 senate election and once more it is clearly an outlier, so we have another indication of fraud.

It is worth noting the complementarity of such methods. 
Shortly, for a fixed base $b$ we are mainly interested in studying the KL divergence for the leading digit frequency when comparing it with the Benford-Newcomb distribution and we are also interested in the ratio of the probabilities $R(d_1,d_2)$. For the KL divergence, we already take into account the frequencies of all $b-1$ possible symbols for the first digit. However, we would have $(b-1)(b-2)/2$ possible combinations of symbols to analyze in order to study all the possible entries of the function $R(d_1,d_2)$. 
o
Since we will vary $b$ from $6$ to $70$, it is not reasonable to analyze all the possible combinations for those ratios for every base we chose. Therefore, we must select which digits $d_1$ and $d_2$ are the most relevant to be studied for the leading digit. We chose $d_1 = 1$ and $d_2=2$ and point out three reasons to do so. 

The first one is obvious since we must choose digits that exist in every base we are considering. The second one is because the frequencies of the leading digit being equal to $1$ or $2$ are good markers to the Benford-Newcomb Law since the theoretical probability of the event the first digit equals to $1$ or $2$ is significantly heavy independently of the chosen base. Indeed, the probability of this event is given by the following function: 

\begin{equation}
    M(b)=\frac{ln(2)}{ln(b)}+\frac{ln(3/2)}{ln(b)}
\end{equation}
where $b$ is an integer greater than $3$. This function for $b$ in the range we are interested in is bounded below by $0.25$, which justifies our claim to analyze that event as a marker.  The third reason is that the probabilities between the first two digits are large enough that it is statistically easier to distinguish between them, as opposed to digits farther ahead.

\subsubsection*{Method 3 - Standard Deviation}

The third employed method allows us to somehow quantify the chance of fraud in the analyzed elections. As we did in the first method,o the function defined in Eq.\ref{Procedimento1} is used. However, we now generate a sample for each election and compute for these data the standard deviation. 

Following that, we build another set with all the values of the standard deviations obtained for each one of the 45 elections considered. Therefore we have a new sample whose histogram is presented in Fig.\ref{grn}.

When one approaches the distribution by a normal function it is possible to estimate how far is the value taken from Bahia-94. This distance is then measured in SD units and it is always beyond 3SD (almost 4 in the presented example). Therefore, from the usually expected distribution, we may argue that the chance of being only a statistical fluctuation, i.e. not an intentional fraud, is less than $0.2\%$.

\begin{figure}[ht]
\epsfig{file=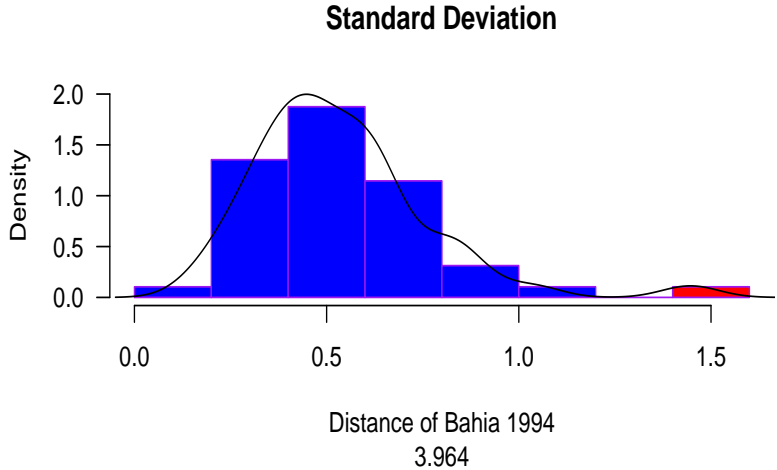,width=11cm,height=7cm}
\caption{Histogram of the Standard Deviation for each election computed as in Figure. \ref{grn}. The distance from 1994-BA SD - small red bar - to the mean is about 3.96$\sigma_{sd}$. ($\sigma_{sd}$ means the standard deviation of the above data)}\label{sd}
\end{figure}

\subsubsection*{Benford $\times$ Uniform}

\begin{figure}[ht]
\epsfig{file=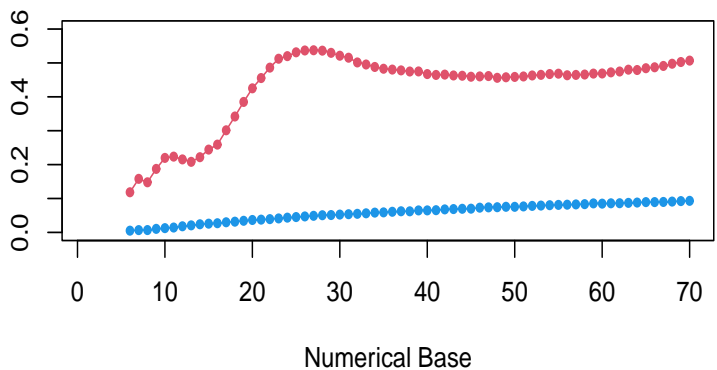,width=11cm,height=7cm}

\caption{Mean KL distance computed for all the considered elections to Uniform Distribution (red line) and Benford-Newcomb Distribution (blue line).}\label{unibenf}
\end{figure}

We also present an additional result in which Kullback–Leibler divergence was computed in order to compare the first-digit measured frequencies to Benford-Newcomb Distribution (BND) and Uniform Distribution (UD) where the frequencieos are assumed to be equal. It is very clear from the behavior of KLD for each distribution as presented in Figure \ref{unibenf} that Benford fits much better the first digit data than Uniform Distribution. 

This result indicates that Benford's distribution must be important in verifying the fairness of an election.

\section*{Conclusion}

Our initial contribution is certainly the use of base invariance of Benford-Newcomb Law. We show that it may help uncover patterns that are not perceptible when applying the Benford-Newcomb Law in the usual way. Although only elections are studied, this method certainly can be employed to look for suspicious data in other contexts.

Some authors have argued that BND is not a good tool when applied to elections~\cite{benfordruim}. We believe we have shown that Benford-Newcomb can be useful as a forensic tool for elections, although noot as the single method to be used. We also must point out that some caution must be taken -- in particular, when trying to detect 'outliers' one has to compare similar elections. For instance, we should not compare Iranian presidential elections with Brazilian ones. 

Considering the particular analyzed event, although too far in the past, this work sheds light on fraud evidence that, so far as we know, has never received a similar (statistical) approach despite the complaints by Waldyr Pires party members. 

We may say finally that such a method could be useful in conjunction with others, such as those described by Mebane and Hicken~\cite{forensic} for election forensics, and perhaps also in other areas.

Finally, it is worth mentioning that our analysis detected no outliers in similar Brazilian elections where electronic ballot boxes were used, a fact which could be seen as supportive of the hypothesis that such ballot boxes do indeed present stronger fraud resistance than older methods (this point is currently being challenged by a number of players in Brazilian politics, giving rise to a dispute between the Brazilian Supreme court and the current Brazilian Secretary of Defense~\cite{tserespondequestionamentos}).

\section*{Acknowledgements}

We would like to thank Isabella Gonçalves de Alvarenga for the helpful suggestions and fruitful discussion.

\end{document}